\documentclass[12pt]{article}
\usepackage[english]{babel}
\usepackage{amssymb}
\usepackage[centertags]{amsmath}
\title{\textbf{Null-curves in $\mathbb{R}^{2,n}$ as flat dynamical systems}}
\author {A. M. Latyshev, S. L. Lyakhovich and A. A. Sharapov\footnote{The work was partially supported by  the RFBR grant No. 13-02-00551. A. A. Sharapov would like to acknowledge support from Dynasty Foundation.}}
\date{\footnotesize\textit{Physics Faculty, Tomsk State University, Tomsk 634050, Russia }}
\begin{document}

\maketitle

\begin{abstract}
 We prove that the differential equation for the null-curves of pseudo-Euclidean space $\mathbb{R}^{2,n}$ defines a flat dynamical system in the sense of optimal control theory. The connection with general gauge theories is briefly discussed.
\end{abstract}

\textbf{0.} A system of first-order ordinary differential equations
\begin{equation}\label{ODE}
f_a(x, x')=0\,,\qquad a=1,\ldots, l\,,
\end{equation}
in the unknowns $x=(x_1,\ldots, x_m)$ is called \textit{flat} if the general solution to the system can be represented in the form
\begin{equation}\label{Sol}
x=g(y, y', \ldots, y^{_{(p)}})\,,
\end{equation}
 where $y=(y_1,\ldots, y_k)$ are arbitrary functions of the independent variable; in this case a vector function $h(x, x', \ldots, x^{(q)})$ should exist such that the substitution
\begin{equation}\label{out}
y=h(x, x',\ldots, {x}^{_{(q)}})
\end{equation}
makes Eq. (\ref{Sol}) into an identity with Eqs. (\ref{ODE}) taken into account. Functions (\ref{out}) are called the \textit{flat outputs} for system (1). If $p$ is the order of the highest derivative of $y$ contained in $g$, then one speaks of a $p$-flat dynamical system, and the number $p$ is called the \textit{order of flat control}. We will consider only regular systems, i.e., systems for which $\mathrm{rank}(\partial f/\partial x')=l$. In this case, the number of functions determining the flat output (\ref{out}) is equal to $k=m-l$. As usual, at generic points we will assume the functions $g$ and $h$ to be real-analytic.

Flat dynamical systems appeared and have been actively studied in
the mathematical theory of optimal control \cite{FLMR} although the
concept of flatness goes back to Hilbert's work \cite{H} on systems
of differential equations that are integrable without quadratures.
Unfortunately, no effective criteria are known to date, allowing one
to decide whether a given dynamical system is flat or not. There is
a close relationship between the controllable dynamical systems
considered in mathematics and the general gauge theories studied in
physics \cite{LSh}. From a physical point of view, the presence of
control is equivalent to gauge invariance of the equations of
motion. In particular, flat dynamical systems can be characterized
as gauge theories without physical degrees of freedom, the finite
(non-infinitesimal) gauge transformations of which are local, i.e.,
they involve a finite number of derivatives of the gauge parameters.
The action of gauge transformations on solutions (\ref{Sol}) is
induced by the translation $y\rightarrow y+\varepsilon$, so that the
gauge group turns out to be abelian. Flat outputs (\ref{out}) are
interpreted in this context as the gauge conditions $h=0$. Note that
the corresponding Fadeev-Popov matrix $(\delta_\varepsilon h/\delta
\varepsilon)$ is given by the unit operator.

\vspace{3mm}{\textbf{1.}} In this note we consider differential equations that describe null-curves in pseudo-Euclidean spaces $\mathbb{R}^{2,n}$. These curves $x(\tau)$ satisfy the equation
\begin{equation}\label{null}
(x', x')=0\,,
\end{equation}
where the round brackets denote the scalar product in $\mathbb{R}^{2,n}$. We use a metric with signature $(--+\cdots+)$. The study of the null-curves in pseudo-Euclidean spaces with different signatures is of interest from both a physical and a mathematical point of view. For example, null-curves of Minkowski space correspond to the world lines of massless particles. The problem of a local description of minimal two-dimensional surfaces (relativistic strings) also reduces to finding a pair of null-curves, see e.g. \cite{B}, \cite{HSh1}, \cite{HSh2}.

Notice that equation (\ref{null}) is invariant with respect to reparametrization $\tau \rightarrow \tau'(\tau)$ of curves. This gives one functional parameter in the general solution. On the other hand, by making use of reparametrization invariance, it is possible to impose one additional condition on the variables $x$ and $y$. We will use this remark below.

The main result of this note is the proof of the following
statement.

\vspace{2mm}\noindent
\textsc{Theorem.} \textit{The equation of null-curves in $\mathbb{R}^{2,n}$ defines a $2$-flat dynamical system for $n\neq 2$ and a $1$-flat system for $n=2$.}

\vspace{2mm}

The proof will be divided into a few steps.

\vspace{3mm}{\textbf{2.}}
Let us begin with the case of three-dimensional Minkowski space $\mathbb{R}^{2,1}$. Following Weierstrass, we consider the space-like curve
\begin{equation}\label{u}
u(\tau)=(1-\tau^2, \;2\tau,\; 1+\tau^2)\,.
\end{equation}
This curve lies entirely within the light cone, i.e., $(u,u)=0$, and, in addition, $(u', u')=-4$. Setting $\mathrm{x}'=ua$, where $a$ is an arbitrary function of $\tau$, the general solution to the equation $(\mathrm{x}',\mathrm{x}')=0$ can be written in the form
$$
\mathrm{x}(\tilde{\tau})=\int_0^{\tilde{\tau}} uad\tau\,.
$$
In order to avoid quadratures, we set $a=f'''$ and integrate by
parts twice in the last expression. Omitting inessential additive
constants, we obtain
\begin{equation}\label{WH}
\mathrm{x}=uf''-u'f'+u''f\,.
\end{equation}
This is the desired parametrization of the set of null-curves in $\mathbb{R}^{2,1}$ by two functional parameters. As was noted above, one of the parameters is obtained by the reparametrization $\tau\rightarrow g(\tau)$, after which the second one is given by the composition $h (\tau)=f(g(\tau))$. Note that solution (6) substantially involves the second derivative of $h$. We call solution (\ref{WH}) a Weierstrass-Hitchin parametrization \cite{Hit}.

\vspace{4mm}\textbf{3.} In the next step, we consider an auxiliary equation defining a set of space-like curves in $\mathbb{R}^{2,1}$ with prescribed infinitesimal length $\Delta(\tau)$:
\begin{equation}\label{d}
(x',x')=-\Delta^2\,.
\end{equation}
It turns out that the equation above also defines a 2-flat dynamical system, whose general solution can be written in the form\footnote{This problem was considered in \cite{G}, where with the help of quite cumbersome differential-geometric constructions it was shown that system (\ref{d}) is $3$-flat. Our construction, being elementary, allows us to lower the order of the flat control by one.}
$$
x=\mathrm{x}+\frac12 u\Delta\,,
$$
where the curves $u(\tau)$ and $\mathrm{x}(\tau)$  are given by expressions (\ref{u}) and (\ref{WH}). The functions describing the flat output have the form
$$
  \tau= \frac{x_2'-\Delta}{x'_1+x'_3}\,,\qquad f=\frac12\left(\frac{x'_2-\Delta}{x_1'+x'_3}\right)^2(x_3+x_1)-\left(\frac{x'_2-\Delta}{x_1'+x'_3}\right)x_2+\frac12(x_3-x_1)\,.
$$
Parametrization of the set of null-curves in $\mathbb{R}^{2,n}$ can now be obtained if we set
\begin{equation}\label{D}
\Delta=\sqrt{x'^2_4+x'^{2}_5+\cdots +x'^2_{n+2}}
 \end{equation}
and assume $x_4$, $x_5$, ..., $x_{n+2}$ to be arbitrary functions of $\tau$. Together with $g$ and $h$ this gives $n+1$ arbitrary functional parameters in the solution, as should be the case for a regular equation in $n+2$ variables.

\vspace{3mm}\textbf{4.} In the case of the space $\mathbb{R}^{2,2}$ one can lower the order of flat control by one, so that the corresponding system of null-curves appears to be $1$-flat. Let us consider a pair of straight lines that are null and mutually orthogonal
$$
  u(\tau)=(1,\tau, 1, -\tau)\,,\qquad v(\tau)=(-\tau, 1, \tau, 1)\,,
$$
$$
(u,u)=0\,,\qquad (u,v)=0\,,\qquad (v,v)=0\,.
$$
The velocity vector can now be represented in the form $x'=uf''+vg''$ for some functions $f$ and $g$. Integrating the last expression by parts as in Sec. 2, we obtain the following representation for the null-curves \cite{B}, \cite{HSh1}:
$$
x=uf'+vg'-uf-vg\,.
$$
The inversion formulas have the form
$$
f =\frac12\left(\frac{x'_2-x'_4}{x'_3+x'_1}\right)(x_3+x_1)-\frac12(x_2-x_4)\,,\qquad
g =\frac12\left(\frac{x'_3-x'_1}{x'_2+x'_4}\right)(x_2+x_4)-\frac12(x_3-x_1)\,,
$$
$$
\tau=\frac{x'_2-x'_4}{x'_3+x'_1}=\frac{x'_3-x'_1}{x'_2+x'_4}\,.
$$

\vspace{3mm}{\textbf{5.}} We will show, finally, that the order of flat control cannot be less than two for $n>2$. The strategy of the proof is based on the following two remarks. If $x=x(y,y')$ is the general solution of the equation (4), then
$$
d=\mathrm{rank} \left(\frac{\partial (x_1,x_2,\ldots, x_{n+2})}{\partial (y_1,\ldots, y_{n+1},  y'_1,\ldots, y'_{n+1})}\right)\geq n+2\,.
$$
In the opposite case, all null-curves of the family $x=x(y,y')$ would lie on some surface of $\mathbb{R}^{2,n}$ of dimension $d<n+2$, which would contradict the fact that it is possible to draw a null-curve through each point of $\mathbb{R}^{2,n}$. The second remark is that the maximum dimension of a null-subspace in $\mathbb{R}^{2,n}$ equals two and, consequently, any three mutually orthogonal null-vectors are of necessity linearly dependent.
It is obvious that the system cannot be $0$-flat since then we would have $x=x(y)$ and, consequently, $d\leq n+1$.

Let us suppose that the system is $1$-flat. In this case $x=x(y,y')$ and
$$
x'=Dx + \sum_{\alpha=1}^{n+1}y''_\alpha x_{(\alpha)}\,, \qquad D=\sum_{\alpha=1}^{n+1}y'_\alpha \frac{\partial }{\partial y_\alpha}\,,\qquad x_{(\alpha)}=\frac{\partial x}{\partial y'_\alpha}\,.
$$
By virtue of the arbitrariness of $y''_\alpha$ it follows from (\ref{null}) that
\begin{equation}\label{diff}
(Dx,Dx)=0\,, \qquad (Dx,  x_{(\alpha)})\qquad (x_{(\alpha)}, x_{(\beta)})=0\,.
\end{equation}
At each generic point the vectors $x_{(\alpha)}$ generate either a one-dimensional or a two-dimensional null-subspace $I$. Let us consider these possibilities separately.

Let $\dim I=1$. Without loss of generality, we can assume that $I$ is generated by a null-vector $x_{(1)}$. Then there exist functions $\lambda_a$, $a=2,3,\ldots, n+1 $, such that
$$
V_ax=0\,,\qquad V_a=\frac{\partial}{\partial y'_a}-\lambda_a\frac{\partial}{\partial y'_1} \,.
$$
Considering the differential consequences of Eqs. (\ref{diff}), we find that
$$
V_a(Dx,Dx)=2(W_ax ,Dx)=0\,,\qquad V_a(Dx, x_{(1)})=(W_a x, x_{(1)}) =0\,,$$
$$
\qquad W_a=\frac{\partial}{\partial y^a}+\lambda_a\frac{\partial}{\partial y_1}\,.
$$
Consequently, the vectors $W_ax$ are orthogonal to the null-subspace $J$ generated by the vectors $Dx$ and $x_{(1)}$. Hence it is easy to see that for any dimensionality $J$ the vectors $W_ax$, $Dx$, and $x_{(1)}$ are linearly dependent, i.e., there exist functions $\xi^a$, $\mu$, and $\lambda$, not simultaneously equal to zero, such that
$$
Wx=0\,,   \qquad  W=\xi^aW_a+\mu D+\lambda\frac{\partial }{\partial y'_1}\,.
$$
The vector fields $V_a$ and $W$ together generate an $(n+1)$-dimensional vector distribution (not necessarily integrable), leaving the functions $x(y,y')$ invariant. Consequently,  $d \leq n+1$.
Let $\dim I=2$. It can be assumed that $I=\mathrm{span}(x_{(1)}, x_{(2)})$. Then it follows from relations (\ref{diff}) that the vectors  $Dx$ and $x_{(A)}$, where $À=3,\ldots, n+1$, belong to $I$. In other words, there exist functions $\sigma$, $\nu$, $\lambda_A$, and $\mu_A$, such that
\begin{equation}\label{DV}
\tilde{D}x=0\,,\qquad V_Ax=0\,,\qquad
\tilde{D}=D+\sigma\frac{\partial }{\partial y'_1}+\nu\frac{\partial }{\partial y'_2}\,,\qquad V_A=\frac{\partial }{\partial y'_A}+\lambda_A\frac{\partial x}{\partial y'_1}+\mu_A\frac{\partial x}{\partial y'_2}\,.
\end{equation}
The commutator of the vector fields $\tilde{D}$ and $V_A$ has the form
$$
\tilde{W}_A=[V_A, \tilde{D}]=\frac{\partial}{\partial y_A}+\lambda_A\frac{\partial}{\partial y_1}+\mu_A\frac{\partial}{\partial y_2}+\cdots\,,
$$
where the dots stand for the terms proportional to $\partial/\partial y'_1$ and $\partial/\partial y'_2$. The equations $\tilde{W}_A x=0$ are the integrability conditions for system (\ref{DV}). It is obvious that the vector fields $V_A$ and $\tilde{W}_A$. are linearly independent and generate a $(2n-2)$-dimensional distribution. The invariance condition for the function $x(y,y')$ with respect to the action of this (not necessarily integrable) distribution implies that $d\leq (2n+2)-(2n-2)=4$. Consequently, in a dimension greater than four the functions $x=x(y,y')$ cannot parametrize all the null-curves. In four dimensions the null-curves in fact constitute a $1$-flat dynamical system, as was shown above.

\textsc{Remark 1.}
To this date, it is unknown whether null-curves of Minkowski space generate a flat dynamical system in a dimension greater than four. With the help of arguments similar to those used in the present work, it is possible to show that the order of flat control for null-curves in $\mathbb{R}^{1,n}$ cannot be less than $n$. This is in agreement with the results of \cite{HSh1}, where with the help of the twistor technique a parametrization of null-curves in $\mathbb{R}^{1,3}$ was proposed involving the third derivative of the functional parameters. Unfortunately, efforts to extend this technique to higher dimensions have still not led to success \cite{HSh2}.

\textsc{Remark 2.} As we saw above, in the case of $\mathbb{R}^{2,1}$ and $\mathbb{R}^{2,2}$ parametrizations of a set of null-curves and the flat outputs are given by rational functions. In higher dimensions the functions of the flat outputs involve radicals. By changing the signature of the quadratic form in Eq. (\ref{D}), it is possible to obtain a parametrization of the null-curves in pseudo-Euclidean spaces $\mathbb{R}^{p,q}$ with $\min (p,q)>2$. The possible velocity vectors of these curves will cover, however, only some open region in the light cone. The existence in the case of higher derivatives of a rational parametrization with rational flat outputs remains an open question.

\textsc{Remark 3.} It is easy to see that parametrization of the null-curves in pseudo-Euclidean spaces simultaneously gives a local representation for null-curves on conformally flat pseudo-Euclidean manifolds, for example, spaces of constant curvature. It would be interesting to describe a class of pseudo-Riemannian metrics allowing a flat control over null-curves of a given order.

\makeatother
\end{document}